\documentclass[conference]{IEEEtran}
\IEEEoverridecommandlockouts
\usepackage{cite}
\usepackage{amsmath,amssymb,amsfonts}
\usepackage{algorithmic}
\usepackage{graphicx}
\usepackage{textcomp}
\usepackage{xcolor}
\usepackage{multirow}
\usepackage[T1]{fontenc}
\usepackage{lmodern}
\usepackage{times}

\def\BibTeX{{\rm B\kern-.05em{\sc i\kern-.025em b}\kern-.08em
    T\kern-.1667em\lower.7ex\hbox{E}\kern-.125emX}}
\begin{document}

\title{Deep Reinforcement Learning for Interference Suppression in RIS-Aided Space–Air–Ground Integrated Networks\\
}

\author{\IEEEauthorblockN{\textsuperscript{1}Pujitha Mamillapalli, \textsuperscript{2,3}Shikhar Verma, \textsuperscript{3}Tiago Koketsu Rodrigues
and \textsuperscript{1}Abhinav Kumar,
\IEEEauthorblockA{\textit{\textsuperscript{1}Department of Artificial Intelligence, Indian Institute of Technology Hyderabad, Telangana, India}}
\IEEEauthorblockA{\textit{\textsuperscript{2} School of Informatics, Kami, Kochi University of Technology, Kochi, Japan,}}
\IEEEauthorblockA{\textit{\textsuperscript{3}Graduate School of Information Sciences, Tohoku University, Sendai, Miyagi, Japan }}}}

\maketitle

\begin{abstract}
Future 6G networks envision ubiquitous connectivity through space-air-ground integrated networks (SAGINs), where high-altitude platform stations (HAPSs) and satellites complement terrestrial systems to provide wide-area, low-latency coverage. However, the rapid growth of terrestrial devices intensifies spectrum sharing between terrestrial and non-terrestrial segments, resulting in severe cross-tier interference. In particular, frequency sharing between the HAPS satellite uplink and HAPS ground downlink improves spectrum efficiency but suffers from interference caused by the HAPS antenna back-lobe. Existing approaches relying on zero-forcing (ZF) codebooks have limited performance under highly dynamic channel conditions. To overcome this limitation, we employ a reconfigurable intelligent surface (RIS)-aided HAPS-based SAGIN framework with a deep deterministic policy gradient (DDPG) algorithm. The proposed DDPG framework optimizes the HAPS beamforming weights to form spatial nulls toward interference sources while maintaining robust links to the desired signals. Simulation results demonstrate that the DDPG framework consistently outperforms conventional ZF beamforming among different RIS configurations, achieving up to \(11.3\%\) throughput improvement for a \(4\times4\) RIS configuration, validating its adaptive capability to enhance spectral efficiency in dynamic HAPS-based SAGINs.
\end{abstract}


\section{Introduction}
In the forthcoming 6G era, wireless communication is expected to extend far beyond terrestrial boundaries, providing seamless connectivity across land, air, underwater, and space~\cite{Liu2018}. Conventional terrestrial networks, constrained by limited coverage, static infrastructure, and line-of-sight (LoS) blockages, cannot support such ubiquitous connectivity. Space–air–ground integrated networks (SAGINs) have thus emerged as a promising architecture, interconnecting satellites, high-altitude platform stations (HAPSs), and ground users in a multi-layered structure to enable continuous, global wireless coverage~\cite{Liu2018}. Within SAGIN, HAPS acts as a critical intermediary between satellites and ground terminals, offering quasi-stationary deployment, wide-area coverage, and favorable LoS conditions~\cite{Alam2021, Jia2023, Ding2022}.

Despite these advantages, efficient communication in HAPS-based SAGINs remains challenging due to inter-link interference. When the HAPS satellite uplink and HAPS ground downlink share the same frequency spectrum, back-lobe radiation from the uplink antenna can interfere with downlink users, degrading signal quality. Traditional interference mitigation techniques, such as frequency-division or time-division multiplexing, provide partial isolation but result in inefficient spectrum utilization~\cite{Matsushita2022}. Advanced spatial processing techniques, including zero-forcing (ZF) beamforming and codebook-based null steering~\cite{kawamoto2024interference}, can suppress interference by shaping transmit beams toward desired directions while forming nulls toward unintended users. However, these approaches are limited by the number of users, discrete codebook resolution, and dynamic network conditions, leading to quantization errors and suboptimal performance in dense or rapidly changing environments.

Reconfigurable intelligent surfaces (RISs) have recently emerged as powerful enablers for intelligent, energy-efficient wireless communication. With passive elements that independently adjust phase shifts, RISs can reshape the propagation environment to enhance desired signals and suppress interference without additional power or spectrum usage~\cite{Li2025b}. Integrating RISs with HAPS platforms enables adaptive beam steering, environment-aware interference control, and improved spectral reuse across space, air, and ground segments~\cite{Ramezani2023}. Due to the complexity of RIS-aided HAPS-based SAGINs, traditional optimization methods or supervised deep learning are often inefficient, as they require full channel knowledge and cannot scale to the exponential RIS configuration space~\cite{Wu2023}. By contrast, deep reinforcement learning (DRL) can learn optimal strategies directly from environmental interactions, making it suitable for dynamic non-terrestrial networks~\cite{Wu2023}.

This work proposes a RIS-aided HAPS-based SAGIN framework to mitigate inter-link interference through null directivity optimization. A DRL-driven DDPG optimization framework~\cite{FNN_arch} is employed to dynamically adjust the HAPS beamforming weights, forming antenna nulls toward interference sources while preserving the required quality-of-service (QoS) for intended users. By exploiting the design of HAPS beamforming, the proposed approach effectively suppresses interference arising from frequency sharing between uplink and downlink transmissions. Overall, it provides a scalable, energy-efficient, and adaptive interference mitigation solution for next-generation 6G non-terrestrial networks.

\textit{Notation}: $a$, $\mathbf{a}$, and $\mathbf{A}$ denote a scalar, column vector, and matrix, respectively. $(\cdot)^T$ and $(\cdot)^H$ denote the transpose and Hermitian transpose, respectively. $\mathbf{A}(k,i)$ is the $(k,i)$ entry of $\mathbf{A}$. $\mathbb{C}^{n \times n}$ denotes the set of $n \times n$ complex matrices. $\mathbf{I}$ is the identity matrix. $\lVert \cdot \rVert$ is the $l_2$-norm. $j$ is the imaginary unit. $\nabla f(x)$ is the Euclidean gradient of $f(x)$ with respect to $x$. $\mathcal{CN}(\mu,\sigma^2)$ denotes a circularly symmetric complex Gaussian random variable with mean $\mu$ and variance $\sigma^2$.
\begin{figure}[t]
\centering
\includegraphics[width=0.9\linewidth]{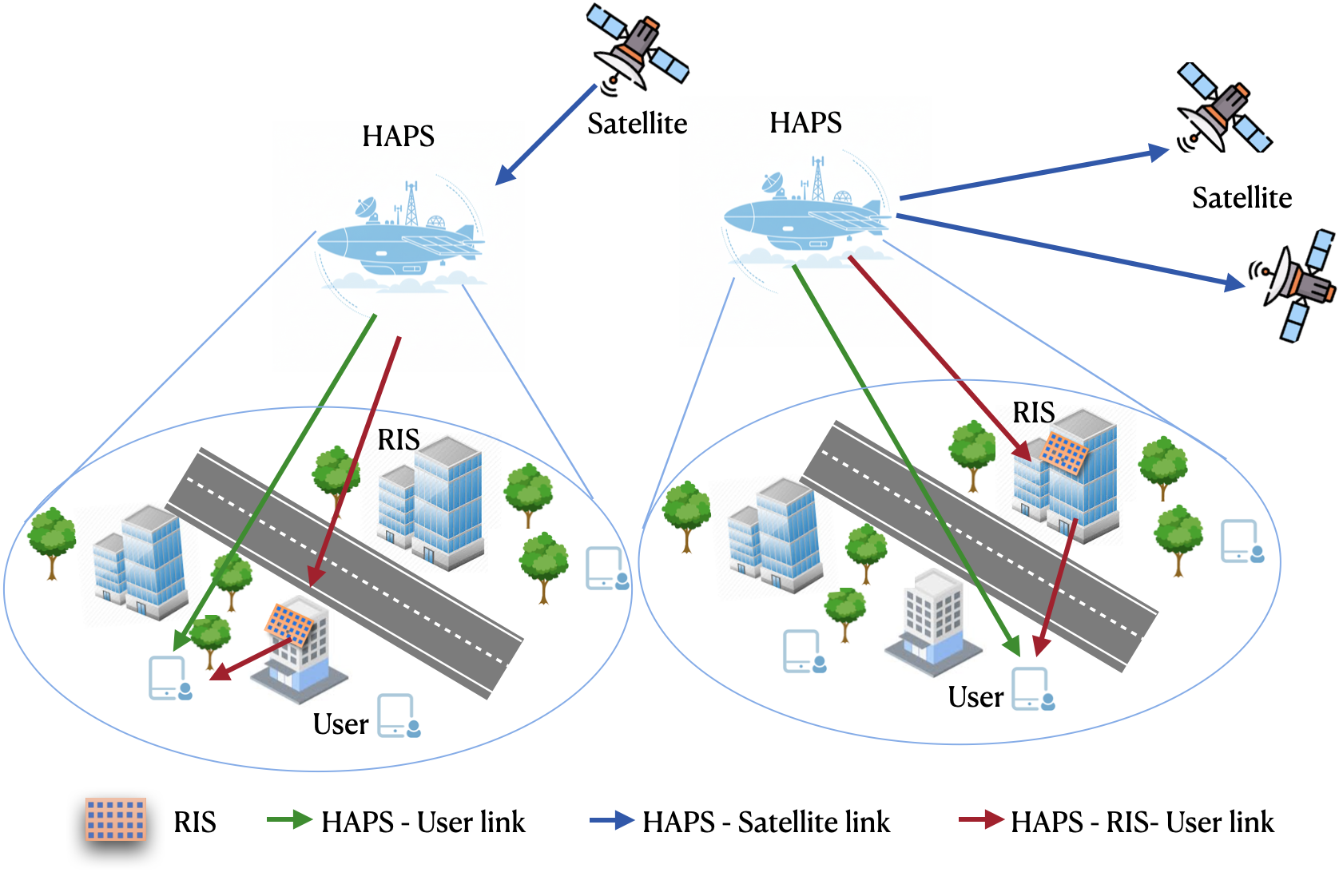}
\caption{Illustration of the RIS-aided HAPS-based SAGIN system model. The figure illustrates the communication links between the satellite, RIS, high-altitude platform station (HAPS), and ground users.}
\label{fig:System_Model}
\vspace{-0.2in}
\end{figure}

\section{System Model}
We consider a RIS-aided HAPS-based SAGIN architecture, as illustrated in Fig.~\ref{fig:System_Model}. The system comprises of three communication layers: (i) a \textit{satellite layer} providing wide-area backhaul connectivity, (ii) an \textit{aerial layer} represented by the HAPS equipped with uniform planar array (UPA) antennas, and (iii) a \textit{ground layer} consisting of a RIS and multiple users within the HAPS coverage area. The HAPS acts as an intermediate relay between the satellite and ground users, simultaneously supporting uplink and downlink transmissions within a shared frequency band. The uplink channel corresponds to the HAPS satellite (HS) link, while the downlink channel represents the HAPS ground (HG) link. 

\subsection{HAPS Antenna Configuration} The HAPS is equipped with two UPA antenna systems, each comprising \( N \) antenna elements, as follows: \begin{enumerate} \item an \textit{HS antenna array} dedicated to uplink transmission toward the satellite, and \item an \textit{HG antenna array} dedicated to downlink transmission toward the ground users. \end{enumerate} Let \( K_{\text{sat}} \) and \( K_{\text{UE}} \) denote the numbers of satellites and ground users, respectively. Each HAPS employs two UPAs, where the uplink (HS) array communicates with \( K_{\text{sat}} \) satellites and the downlink (HG) array serves \( K_{\text{UE}} \) users. Both arrays operate in the same frequency band to maximize spectral efficiency. However, due to their close placement, back-lobe radiation from the uplink array couples into the co-located downlink array, causing \textit{inter-link interference}. This asymmetry arises because the uplink array operates as a high-gain transmitter whose side-lobe and back-lobe emissions affect the downlink receiver. In contrast, the downlink beam is directed toward the ground, and its upward leakage is negligible~\cite{ITU_R_F2437_2018}.

\subsection{RIS-aided Propagation Environment}
To mitigate interference from the HAPS uplink transmission and enhance the received signal quality, a RIS with $L \times L$ reflecting elements is deployed within the HAPS coverage region. Each RIS element introduces a controllable phase shift to the incident signal, thereby shaping the propagation environment intelligently. The reflection matrix of the RIS is defined as
\begin{equation}
\mathbf{\Theta} = \mathrm{diag}\left(e^{j p_{1}}, e^{j p_{2}}, \ldots, e^{j p_{l}}, \ldots, e^{j p_{L^{2}}}\right),
\label{eq:RIS}
\end{equation}
where \( p_l \in [0, \pi] \) denotes the adjustable phase shift applied by the \( l^{\text{th}} \) reflecting element. When an electromagnetic wave impinges on the RIS with an incidence angle \( \omega_{\text{in}} \) and is reflected toward a desired direction \( \omega_{\text{out}} \), the phase shift of each element must be configured to ensure constructive interference in the target direction. The required phase configuration for the \( l^{\text{th}} \) element is expressed as in \cite{KawamotoRIS_phase}
\begin{equation}
p_{l} = -\frac{2\pi f d_{\text{RIS}} l}{c} \left( \sin{\omega_{\text{in}}} + \sin{\omega_{\text{out}}} \right),
\label{eq:RIS_phase}
\end{equation}
where \( f \) is the carrier frequency, \( c \) denotes the speed of light, and \( d_{\text{RIS}} \) represents the element spacing of the RIS array. This configuration ensures that the reflected signals are coherently combined in the desired direction, enhancing the received signal strength and mitigating interference.

The effective downlink channel from a HAPS to the $k^{th}$ user via the RIS can be modeled as
\begin{equation}
    \mathbf{h}_{\text{DL},k}^H = \mathbf{g}_{k}^H \mathbf{\Theta }\mathbf{H}_{h} + \mathbf{h}_{h,k}^H,
    \label{eq:DL_channel}
\end{equation}

where $\mathbf{H}_{h} \in \mathbb{C}^{L^2 \times N}$ denotes the HAPS–RIS channel, $\mathbf{g}_{k} \in \mathbb{C}^{L^2 \times 1}$ represents the RIS–user channel, $\mathbf{h}_{h,k} \in \mathbb{C}^{N \times 1}$ denotes the direct HAPS–user channel, and $\mathbf{\Theta}$ is the RIS reflection matrix as defined in~\eqref{eq:RIS}.


\subsection{Null Directivity-Based Interference Suppression}
To effectively suppress inter-link interference in the RIS-aided HAPS-based SAGIN, the HAPS transmit signals are collectively modeled using a beamforming matrix defined as
\begin{align}
    \mathbf{W} &= [\mathbf{w}_{1}^H, \cdots, \mathbf{w}_{K_{\text{sat}}}^H, 
    \mathbf{w}_{K_{\text{sat}}+1}^H, \dots, \mathbf{w}_{K_{\text{sat}}+K_{\text{UE}}}^H] \nonumber\\
    &\in \mathbb{C}^{N \times (K_{\text{sat}} + K_{\text{UE}})},
\end{align}
where \( \mathbf{w}_{1}^H, \cdots, \mathbf{w}_{K_{\text{sat}}}^H \) correspond to the beamforming vectors for uplink transmission toward the satellite, while \( \mathbf{w}_{K_{\text{sat}}+1}^H, \dots, \mathbf{w}_{K_{\text{sat}}+K_{\text{UE}}}^H \) represent the downlink beamforming vectors for the \( K_{\text{UE}} \) users.
The received signal vector at the HAPS output, accounting for both the satellite uplink and user downlink transmissions, is expressed as follows, as in~\cite{kawamoto2024interference} 
\begin{equation}
    \mathbf{y}_{\text{HAPS}} = \mathbf{H}_{\text{HAPS}} \mathbf{W} \mathbf{x} + \mathbf{n},
\end{equation}
where \( \mathbf{y}_{\text{HAPS}} \in \mathbb{C}^{(K_{\text{sat}} + K_{\text{UE}}) \times 1} \) represents the received signal vector, in which the first \( K_{\text{sat}} \) elements correspond to the satellite uplink, while the remaining \( K_{\text{UE}} \) elements correspond to the signals toward ground users. The baseband symbol vector is given by \( \mathbf{x} = [x_{1}, \cdots, x_{K_{\text{sat}}}, x_{K_{\text{sat}}+1}, \cdots, x_{K_{\text{sat}}+K_\text{UE}}]^H \). An effective composite channel matrix \( \mathbf{H}_{\text{HAPS}} \in \mathbb{C}^{(K_{\text{sat}} + K_{\text{UE}}) \times N} \) is defined as
\begin{equation}
    \mathbf{H}_{\text{HAPS}} = 
    [\mathbf{h}_{\text{UL},1}^H, \cdots, \mathbf{h}_{\text{UL},K_{\text{sat}}}^H, 
    \mathbf{h}_{\text{DL},1}^H, \cdots, \mathbf{h}_{\text{DL},K_{\text{UE}}}^H]^H,
    \label{eq:HAPS_channel}
\end{equation}
where \( \mathbf{h}_{\text{UL}} \) represents the uplink channel between the HAPS and the satellites, and \( \mathbf{h}_{\text{DL}} \) denotes the effective downlink channel between the HAPS and users, which incorporates the RIS reflection model described in~\eqref{eq:DL_channel}. The noise term \( \mathbf{n} \in \mathbb{C}^{(K_{\text{sat}} + K_{\text{UE}}) \times 1} \sim \mathcal{CN}(\mathbf{0}, \sigma^2 \mathbf{I}) \) denotes the additive white Gaussian noise (AWGN) with noise variance \( \sigma^2 \). The signal-to-interference-plus-noise ratio (SINR) at  $k^{th}$ ground user is then given by
\begin{equation}
    \gamma_k = 
    \frac{\lVert\mathbf{H}_{\text{HAPS}}(k) \mathbf{w}_kx_{k}\rVert^2}{
    \sum_{i \neq k} \lVert\mathbf{H}_{\text{HAPS}}(i) \mathbf{w}_i x_{i}\rVert^2  + \sigma^2 },~\forall k \in \{1, \cdots, K_\text{UE}\},
\end{equation}
\section{Problem Formulation}
A key challenge in the RIS-aided HAPS-based SAGIN is the \textit{inter-link interference} between the uplink and downlink transmissions. The back-lobe radiation of the HAPS uplink (HS) antenna can overlap with the downlink coverage area, introducing unwanted interference to ground users and degrading the overall received signal quality. The objective is to design the beamforming matrix $\mathbf{W}$ to minimize transit power that allows zero interference while satisfying a minimum SINR requirement $\gamma_{\text{min}}$ for all ground users. This can be formulated as an optimization problem as follows.
\begin{align}
\mathcal{P}_1: \quad
& \underset{\mathbf{w}_i}{\text{minimize}} \sum_{i=1}^{K_\text{sat}+K_\text{UE}}\|\mathbf{w}_i\|^2\\
& \text{s.t.}~~(\mathbf{H}_{\text{HAPS}} \setminus \mathbf{H}_{\text{HAPS}}(i)) \mathbf{w}_i=0, \nonumber\\
& ~\forall i \in \{1,\cdots, K_\text{sat}, K_\text{sat}+1,\cdots, K_\text{sat}+K_\text{UE}\},~\label{P1:constraint1} \\
& \frac{\lVert\mathbf{H}_{\text{HAPS}}(k) \mathbf{w}_k x_k\rVert^2}{
    \sum_{i \neq k} \lVert\mathbf{H}_{\text{HAPS}}(i) \mathbf{w}_i x_i\rVert^2  + \sigma^2 } \geq \gamma_{\text{min}}, \nonumber\\
& ~ \forall k \in \{1,\cdots,K_\text{UE}\},\label{P1:constraint2} \\
&\sum_{i=1}^{K_\text{sat}+K_\text{UE}}\|\mathbf{w}_i x_{i}\|^2 \leq P_t \label{P1:constraint3},
\end{align}
The optimization problem $\mathcal{P}_1$ aims to \emph{minimize the total transmit power}, which is motivated by energy efficiency, hardware limitations, and regulatory constraints, while satisfying three key constraints that ensure system functionality and QoS. Constraint \eqref{P1:constraint1} enforces ZF interference nulling by ensuring that the beamforming vector for each receiver lies in the null space of all other users' channels. Here, $(\mathbf{H}_{\text{HAPS}} \setminus \mathbf{H}_{\text{HAPS}}(i))$ denotes the HAPS channel matrix with the $i^{\text{th}}$ row removed; by excluding this row, the intended receiver's channel is preserved while interference to all other receivers is completely suppressed. Constraint \eqref{P1:constraint2} guarantees a minimum SINR $\gamma_{\text{min}}$ for all ground users to maintain QoS. Finally, constraint \eqref{P1:constraint3} restricts the total transmit power to the budget $P_t$, satisfying hardware and regulatory limits. Feasibility of $\mathcal{P}1$ generally depends on system parameters: strict ZF requires sufficient transmit degrees of freedom ($N \ge K_{\text{sat}} + K_{\text{UE}}$), while SINR satisfaction is contingent on whether the required transmit power lies within $P_t$.

The optimization problem $\mathcal{P}_1$ is non-convex and involves continuous beamforming vectors with coupled constraints under time-varying channels. Conventional iterative solvers are computationally expensive, slow to converge, and prone to local minima while requiring recomputation for each new channel realization. Traditional deep learning approaches inherit this computational burden, require labeled data, and struggle to generalize to dynamic channel conditions; they also cannot reliably enforce hard constraints such as ZF and SINR thresholds~\cite{Zhang2019}. DDPG overcomes these limitations with a model-free continuous-action framework that learns directly from the environmental feedback~\cite{FNN_arch}. It handles beamforming vectors, integrates constraints through reward shaping, adapts to dynamic channels in real time, and scales efficiently to optimize transmit power while maintaining QoS in RIS-aided HAPS-based SAGIN systems.

\begin{figure}[t]
\centering
\includegraphics[width=\linewidth]{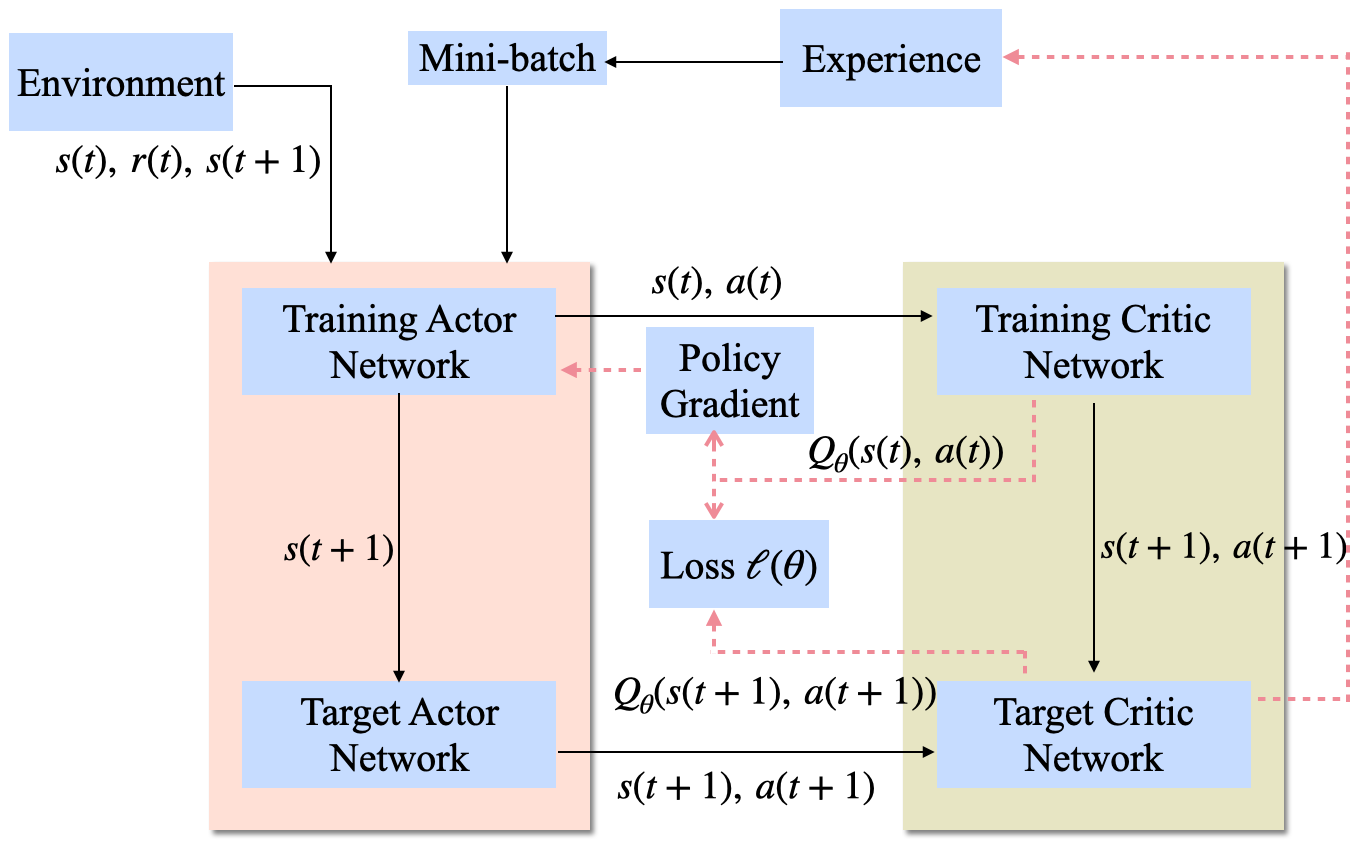}
\caption{Architecture of the DDPG framework: Red dotted arrows indicates  updation; black solid arrows indicates forward‑pass data flow.}
\label{fig:SoftDDPG}
\vspace{-0.1in}
\end{figure}

\section{Deep Deterministic Policy Gradient (DDPG) Framework}
This section employs a DRL framework to optimize the beamforming matrix $\mathbf{W}$ in the RIS-aided HAPS-based SAGIN system. Specifically, the DDPG algorithm is adopted because it can handle continuous action spaces, as illustrated in Fig.~\ref {fig:SoftDDPG}. DDPG is an actor–critic-based DRL approach that integrates two neural networks: the \textit{actor network}, which maps the system state to a continuous action, and the \textit{critic network}, which evaluates the quality of the selected action by estimating the expected cumulative reward. At each time step, the agent interacts with the environment by observing the current system state, executing an action, and receiving a reward signal that reflects the system's performance.

\noindent \textbf{State:} The state represents a set of observations characterizing the environment. At each time step $t$, the system state $s(t)$ captures the channel state information (CSI) of the RIS-aided HAPS-based SAGIN, represented by the composite channel matrix $\mathbf{H}_{\text{HAPS}}$. Since the CSI is complex-valued, its real and imaginary components are separated to construct a real-valued state vector suitable for neural network processing~\cite{FNN_arch}. The impact of imperfect CSI estimation is not addressed in this study and is left for future work.

\noindent \textbf{Action:} The action denotes the decision taken by the agent at time step $t$. In this framework, the action $a(t)$ corresponds to the beamforming matrix $\mathbf{W}(t)$ of the HAPS antenna array. Similar to the state, the real and imaginary parts of $\mathbf{W}(t)$ are separated to ensure compatibility with the deep learning model~\cite{Wu2023}. The action determines how the antenna beams are directed toward the satellites and the ground users to minimize interference and satisfy SINR constraints.

\noindent \textbf{Reward:} The reward $r(t)$ provides feedback to evaluate the effectiveness of the chosen action $a(t)$ given the state $s(t)$. It quantifies system performance in terms of transmit power efficiency, minimum SINR satisfaction, and interference suppression. When the action leads to reduced interference and improved communication quality, a higher reward is assigned, while penalties are applied when constraints are violated.

\noindent \textbf{Policy:} The policy $\rho(s(t), a(t))$ defines the mapping between the observed state and the selected action. It represents the probability of choosing action $a(t)$ for a given state $s(t)$. Training aims to find an optimal policy $\rho^*$ that maximizes the expected cumulative reward over time~\cite{Wu2023}.

\noindent \textbf{State–Action Value Function:} The value function $Q(s(t), a(t))$ measures the expected future cumulative reward obtained by taking action $a(t)$ in state $s(t)$ and following policy $\rho$ thereafter~\cite{FNN_arch}. While the reward provides immediate performance feedback, the value function estimates the long-term benefit of an action, guiding the policy toward optimal decisions.

\noindent \textbf{Experience:} Each agent–environment interaction generates an experience tuple $(s(t), a(t), r(t+1), s(t+1))$, which is stored in a replay buffer. Random sampling from this buffer during training improves learning stability and prevents sample correlation. The Q-value function is parameterized by a neural network with parameters $\theta$:
\begin{equation}
    Q(s(t), a(t)) \triangleq Q_\theta(s(t), a(t)),
\end{equation}
where $\theta$ denotes the network weights and biases. Instead of directly applying the Bellman update as in classical Q-learning~\cite{ICML2018}, DDPG updates $\theta$ using stochastic gradient descent:
\begin{equation}
    \theta(t+1) = \theta(t) - \mu \, \nabla_\theta \ell(\theta),
\end{equation}
where $\mu$ is the learning rate and $\nabla_\theta \ell(\theta)$ is the gradient of the loss with respect to $\theta$.

\subsection{Loss Function}
The loss function measures the mean squared error between the predicted and target Q-values:
\begin{equation}
    \ell(\theta) = \frac{1}{B} \sum_{t=1}^{B} \Big( y(t) - Q_\theta(s(t),a(t)) \Big)^2,
\end{equation}
where $B$ denotes the mini-batch size. The target Q-value $y(t)$ is computed using the target networks:
\begin{equation}
    y(t) = r(t+1) + \tau \max_{a(t+1)} Q_\theta(s(t+1), a(t+1)),
\end{equation}
where $\tau$ is the discount factor controlling the weight of future rewards. The DDPG framework, shown in Fig.~\ref{fig:SoftDDPG}, employs four neural networks to enable stable learning and convergence~\cite{FNN_arch}:
\begin{enumerate}
    \item \textbf{Training Actor Network:} Generates beamforming actions based on the current state.
    \item \textbf{Training Critic Network:} Evaluates the Q-value $Q_{\theta}(s(t), a(t))$ for the current state–action pair.
    \item \textbf{Target Actor Network:} Produces the next action $a(t+1)$ for the updated state $s(t+1)$.
    \item \textbf{Target Critic Network:} Estimates the maximum target Q-value $Q_{\theta}(s(t+1), a(t+1))$ for stable and smooth parameter updates.
\end{enumerate}
The proposed DDPG framework enables the RIS-aided HAPS–based SAGIN system to adapt efficiently to dynamic environments, form null-directed beams, and suppress interference caused by frequency sharing between uplink and downlink links. This approach enhances spectral efficiency, spatial isolation, and energy utilization in SAGIN. Next, the numerical results and performance of the proposed DDPG-based architecture are presented.

\begin{table}[t]
\caption{Simulation parameters}
\label{tab:sim_params}
\centering
\begin{tabular}{l l}
\hline
\textbf{Parameter} & \textbf{Value} \\
\hline
Area width (x-axis) & $100\ \text{m}$ \\
Area width (y-axis) & $100\ \text{m}$ \\
Carrier frequency $f$ & $28\ \text{GHz}$ \\
Speed of light $c$ & $3\times 10^{8}\ \text{m/s}$ \\
Number of Satellites $K_{\text{sat}}$ & $1$ \\
Number of HAPS & $1$ \\
Number of RIS & $1$ \\
Satellite height & $3.2\times 10^{6}\ \text{m}$ \\
HAPS height & $2.0\times 10^{4}\ \text{m}$ \\
Ground user density & $150~\text{users}\text{/km}^{2}$  \\
HAPS antenna configuration $N$ & $50$ elements  \\
AoA at satellite from HAPS  $\upsilon_{\text{sat}}$ & $\pi/4$ \\
RIS elements (x-axis) & $4$ \\
RIS elements (y-axis) & $4$ \\
RIS element spacing $d_{\text{RIS}}$ & $\frac{c}{2f}\approx 5.357\times 10^{-3}\ \text{m}$ \\
Total number of RIS elements $L^2$ & $16$ \\
Transmit power  $P_t$ & 30~dB \\
Bandwidth & 400~MHz\\
\hline
\vspace{-0.2in}
\end{tabular}
\end{table}

\section{Numerical Results and Analysis}
\begin{figure}[t]
\centering
\includegraphics[width=0.9\linewidth]{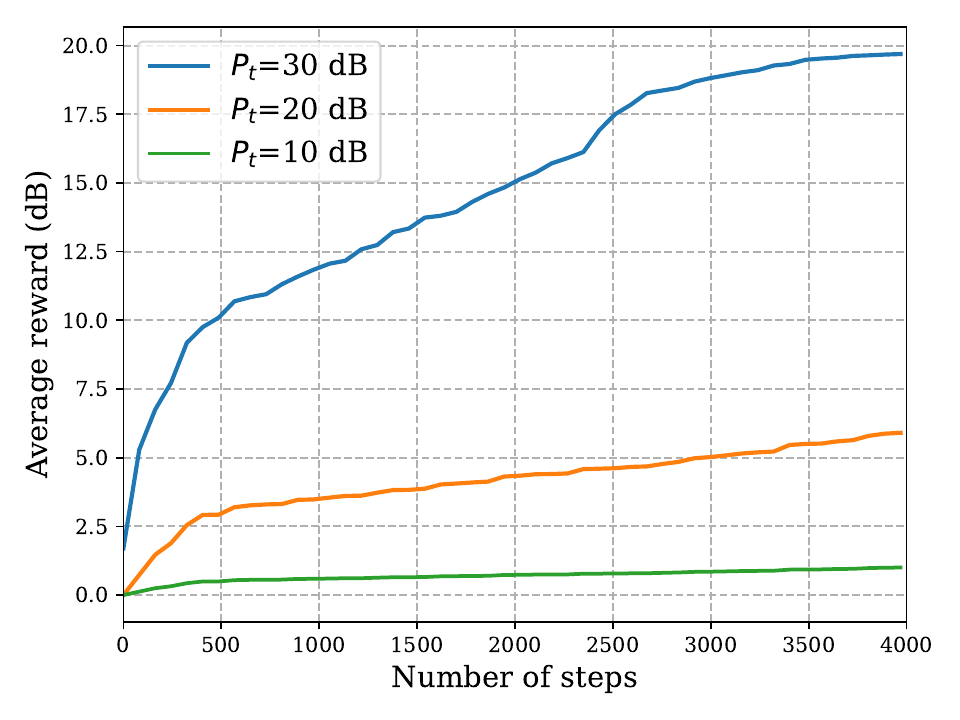}
\caption{Average rewards versus time steps under various transit power $P_t$ with $\gamma_\text{min}=0~\text{dB}$}
\label{fig:reward_fnn}
\vspace{-0.2in}
\end{figure}

This section presents the simulation setup for a single RIS-aided HAPS-based SAGIN framework, as illustrated in Fig.~\ref{fig:System_Model}. The framework is implemented in Python and executed on a workstation equipped with an Intel i9-13900K CPU and an NVIDIA RTX 2080 Ti GPU (12 GB). The system covers a $100 \times 100~\text{m}^2$ ground area with one satellite, a single HAPS, and a RIS supporting communication with multiple ground users randomly distributed according to a Poisson point process with density $150~\text{users/km}^2$. The HAPS operates at 28~GHz using UPAs of 50 antennas for both uplink and downlink, spaced at half-wavelength intervals. The RIS comprises $4 \times 4$ passive reflective elements, each spaced by $c/(2f)$. The satellite and HAPS are located at altitudes of 3200~km and 20~km, respectively.

The wireless channels among the satellite, HAPS, RIS, and users are modeled using a generalized array response of the form $\left[e^{-j\pi (n-1)\sin(\upsilon_{\text{rx}})}\right]_{n=1}^{N}$, where $\upsilon_{\text{rx}}$ denotes the angle of arrival (AoA) and $N$ represents the number of HAPS antennas. The AoA at the satellite is set to $\pi/4$, capturing oblique incidence in the HAPS satellite links. The baseband signal is considered to have unit energy. We assume a minimum SINR requirement of $\gamma_\text{min}=0~\text{dB}$, while the noise variance $\sigma^2$ follows the specifications defined in the 3GPP standards~\cite{3GPP}. This model effectively captures phase variations induced by antenna geometry and spatial directionality—critical factors for mmWave propagation~\cite{kawamoto2024interference}. Key simulation parameters are summarized in Table~\ref{tab:sim_params}.

To ensure computational feasibility and stable convergence of the proposed DDPG framework, a reduced-scale configuration with a limited number of users and RIS elements is adopted, considering the limitations of the available computational hardware. This setup allows the DDPG agent to efficiently explore the action space and learn meaningful beamforming policies without being overwhelmed by excessive dimensionality or interference complexity. Controlling the environment enables systematic assessment of the agent’s capacity to adapt to dynamic channels, mitigate inter-user and inter-link interference, and optimize the trade-off between SINR and transmit power. Furthermore, this controlled configuration demonstrates that the learned beamforming strategies are both theoretically sound and practically implementable, providing a reliable scaling approach to larger RIS-aided HAPS-based SAGIN deployments.

\begin{table}[t]
\centering
\caption{Comparison of Throughput ($\times 10^7$ bps) for Different RIS Configurations and $\alpha$ Values}
\label{tab:throughput_comparison_alpha}
\begin{tabular}{lcccc}
\hline
\textbf{Scheme} & \multicolumn{2}{c}{\textbf{4 $\times$ 4}} & \multicolumn{2}{c}{\textbf{6 $\times$ 6}} \\
\cline{2-5}
 & $\alpha=1$ & $\alpha=2$ & $\alpha=1$ & $\alpha=2$ \\
 
\hline

\textbf{Proposed DDPG} 
& \textbf{4.293} & \textbf{4.048} & \textbf{4.354} & \textbf{4.108} \\

{Zero-Forcing (ZF)~\cite{kawamoto2024interference}} 
& 3.856 & 3.647 & 4.021 & 3.748 \\
\hline
\vspace{-0.2in}
\end{tabular}
\end{table}


The proposed DRL-based interference suppression framework utilizes the DDPG algorithm, where both the actor and critic are designed as fully connected neural networks (FNNs) with two hidden layers~\cite{FNN_arch}. The actor maps the state vector to the continuous action space, with each hidden layer incorporating batch normalization and parametric rectified linear unit (PReLU) activation, and the output scaled to the maximum allowable action. The critic estimates the Q-value by processing the state input through a fully connected layer with batch normalization and PReLU activation, followed by fusion with the action vector in the second layer, and outputs the scalar expected return.

Each training session consists of up to 4,000 steps per episode and a maximum of 10 episodes. A replay buffer of size 1,000 stores past transitions, and mini-batches of size 16 are sampled for updates. A 50-step warm-up phase and Gaussian exploration noise with a standard deviation of 0.1 help stabilize the learning process. The discount factor, learning rate, and weight decay are set to 0.99, 0.01, and 0.00001, respectively. All simulations are initialized with a random seed of 42 for reproducibility. This configuration enables actor and critic networks to effectively capture complex state–action relationships while maintaining stability and efficiency in continuous control environments.

An empirical evaluation of FNN depth reveals that a two-layer hidden architecture achieves a balance between expressivity and generalization, yielding the highest sum rate of 2.09. A single hidden layer attains only 1.821, as it lacks the nonlinearity needed to model the composite HAPS–RIS channel interactions. Extending the depth to three layers yields only a marginal gain of 0.03 over the two-layer hidden model. In contrast, deeper architectures (four or more layers) saturate at around 1.86, indicating over-parameterization, slower convergence, and an increased tendency to overfit. Hence, two (and marginally three) layers offer sufficient hierarchical feature extraction to capture inter-user interference and nonlinear beamforming relationships. Within the computational constraints of Table~\ref{tab:sim_params}, this two-layer design provides optimal representational power, stable convergence, and efficient training performance.

Fig.~\ref{fig:reward_fnn} illustrates the convergence characteristics of the proposed DDPG framework at different levels of transmit power. A performance gradient is observed in terms of power: higher transmit power enables faster convergence and higher asymptotic rewards, reflecting the agent’s improved capacity to explore the state-action space and learn efficient beamforming policies. In contrast, when the transmit power is limited, the agent’s learning process becomes slower and the steady-state reward decreases, as the reduced SINR constrains the achievable throughput and the effectiveness of the learned policy. These findings highlight a positive correlation between transmit power and DDPG performance, confirming the agent’s robust ability to optimize system efficiency under diverse power constraints. 


\begin{figure}[t]
\centering
\includegraphics[width=0.9\linewidth]{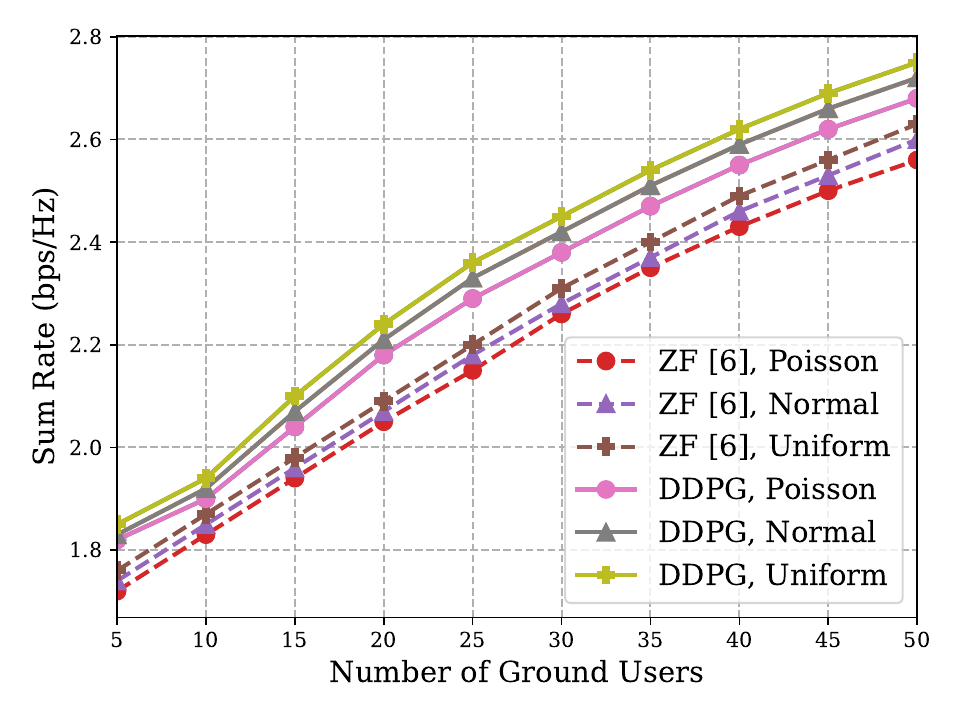}
\caption{Sum rate versus number of ground users for Poisson, Normal, and Uniform user distributions under both DDPG-based and ZF beamforming schemes~\cite{kawamoto2024interference} for a $4\times4$ RIS configuration in the HAPS-based SAGIN framework.}
\vspace{-0.2in}
\label{fig:num}
\end{figure}

Table~\ref{tab:throughput_comparison_alpha} presents throughput comparisons between the proposed DDPG-based RIS-aided scheme and conventional ZF beamforming~\cite{kawamoto2024interference} across various RIS configurations and fairness parameter values ($\alpha$) in the throughput formulation~\cite{fairness}. The parameter $\alpha$ represents the fairness weighting factor that balances total system throughput and user fairness, with a higher $\alpha$ emphasizing more equitable rate distribution among users~\cite{fairness}. For both $4\times4$ and $6\times6$ RIS configurations, the DDPG approach consistently achieves higher throughput, yielding gains of up to $11.34\%$ for $\alpha=1$ and $10.99\%$ for $\alpha=2$. Throughput improvements with larger RIS sizes remain moderate, suggesting diminishing returns beyond a certain number of elements. These results confirm that the DDPG agent effectively leverages RIS-aided beamforming to enhance system throughput under fairness constraints.

Fig.~\ref{fig:num} compares the sum-rate performance of the DDPG-based and ZF beamforming~\cite {kawamoto2024interference} under different user distributions (Poisson, Normal, and Uniform) for a $4\times4$ RIS. In all cases, the sum rate increases with the number of users, and DDPG consistently outperforms ZF beamforming~\cite{kawamoto2024interference}. The performance gap widens with user count, underscoring the DDPG agent’s ability to dynamically optimize the HAPS beamforming matrix for interference mitigation and energy-efficient transmission in multi-user environments. The numerical analysis confirms that the proposed FNN-based DDPG interference suppression framework achieves superior spectral and energy efficiency compared to ZF beamforming~\cite{kawamoto2024interference}. The proposed design presents a scalable, adaptive, and energy-efficient solution for next-generation 6G non-terrestrial networks by optimizing the HAPS beamforming matrix and exploiting RIS-aided channel diversity.

\section{Conclusion}
This paper presented a DDPG framework to address inter-link interference suppression in RIS-aided HAPS-based SAGINs under shared-spectrum conditions. The problem was formulated to minimize total transmit power while satisfying SINR constraints for all the ground users. By leveraging the DDPG algorithm, the system learned to optimize the HAPS beamforming matrix, generating null directivity patterns to suppress interference from the HAPS uplink back-lobe. Numerical results demonstrated that the proposed DRL-based approach outperforms conventional ZF beamforming~\cite{kawamoto2024interference}, achieving up to $11.3\%$ throughput improvement for a $4 \times 4$ RIS configuration. This hybrid RIS-HAPS design offers an energy-efficient and adaptive solution that enhances spectral efficiency and spatial isolation in dynamic 6G non-terrestrial networks. Future work will focus on extending the framework to larger-scale deployments with more users and RIS elements to enhance spatial diversity, network robustness, and adaptability under dynamic user mobility. Moreover, practical considerations such as hardware impairments and imperfect CSI and RIS modeling will be integrated to enhance the framework’s real-world applicability further.



\bibliographystyle{IEEEtran}
\bibliography{IEEEabrv,citation_SAGIN}

\end{document}